\documentclass{article}

    \usepackage[preprint]{neurips_2025}

\usepackage[utf8]{inputenc} 
\usepackage[T1]{fontenc}    
\usepackage{hyperref}       
\usepackage{url}            
\usepackage{booktabs}       
\usepackage{amsfonts}       
\usepackage{nicefrac}       
\usepackage{microtype}      
\usepackage{xcolor} 

\usepackage{graphicx}
\usepackage{amsmath}

\title{Computationally Efficient Information-Driven Optical Design with Interchanging Optimization}

\author{%
  Eric Markley \\
  \And
  Henry Pinkard
  \And
  Leyla Kabuli \\
  \And
  Nalini Singh\\
  \And
  Laura Waller \\
  \AND
  {\normalfont\mdseries\small Department of Electrical Engineering and Computer Sciences, UC Berkeley}
}

\begin{document}

\maketitle
\begin{abstract}

Recent work has demonstrated that imaging systems can be evaluated through the information content of their measurements alone, enabling application-agnostic optical design that avoids computational decoding challenges. Information-Driven Encoder Analysis Learning (IDEAL) was proposed to automate this process through gradient-based optimization. In this work, we study IDEAL across diverse imaging systems and find that it suffers from high memory usage, long runtimes, and a potentially mismatched objective function due to end-to-end differentiability requirements. We introduce IDEAL with Interchanging Optimization (IDEAL-IO), a method that decouples density estimation from optical parameter optimization by alternating between fitting models to current measurements and updating optical parameters using fixed models for information estimation. This approach reduces runtime and memory usage by up to 6× while enabling more expressive density models that guide optimization toward superior designs. We validate our method on diffractive optics, lensless imaging, and snapshot 3D microscopy applications, establishing information-theoretic optimization as a practical, scalable strategy for real-world imaging system design.

\end{abstract}

\section{Introduction}

In computational imaging, optics and algorithms are designed jointly, unlocking novel trade-offs between hardware complexity, system performance, and cost~\cite{bhandariComputationalImaging2022}. A typical system contains an optical encoder that captures a measurement and a software decoder that reconstructs or interprets the scene. With the freedom to implement diverse optical encoding strategies made possible by computational decoding, the optimal encoder design is an open question in many applications.

Traditional approaches to encoder design rely on expert intuition and idealized physical models, often producing suboptimal designs that break down under real-world noise or model mismatch. More recently, end-to-end methods jointly optimize optical parameters and reconstruction algorithms using deep learning~\cite{lin_end--end_2020, sitzmann_end--end_2018, sun_end--end_2021}. These systems treat the image formation model as a differentiable layer in a neural network and train the optics and decoder simultaneously. While this often yields high-performance designs, it requires substantial computational resources~\cite{deb_programmable_2021} and presents practical challenges when optimizing through state-of-the-art reconstruction algorithms~\cite{yang_image_2023}. 

Recent work has suggested that encoders could be designed independently using information-theoretic principles~\cite{pinkard2024informationdrivendesignimagingsystems}. This approach is appealing because it decouples the encoder optimization from specific decoder implementations, focusing instead on measurement quality by maximizing the information content of measurements. Furthermore, information-theoretic metrics naturally balance resolution, signal-to-noise ratio, and other performance factors within a unified framework, providing an application-agnostic approach to optical design.

Information-Driven Encoder Analysis Learning (IDEAL) proposes automating this process through gradient-based optimization of mutual information between scenes and measurements. This approach offers compelling potential advantages: it avoids the challenges of designing and optimizing through decoders such as complex reconstruction algorithms, and as a result can readily be applied to a wide variety of systems. However, the limitations of IDEAL's proposed implementation and effectiveness across diverse imaging systems have not been thoroughly explored.

In this work, we identify a critical limitation of IDEAL in its original formulation: the requirement for end-to-end differentiability through both density estimation and optical parameter optimization introduces prohibitive computational demands and potentially misaligned optimization objectives.

To address these issues, we introduce IDEAL with Interchanging Optimization (IDEAL-IO). Our key insight is that decoupling density model fitting from optical parameter updates enables significantly more efficient optimization while maintaining or improving design quality. We alternate between: (1) fitting a density model to fixed measurements from the current optical system and (2) updating the optics using a differentiable MI estimator while holding the fitted model constant. This eliminates the need for the fitting process itself to be differentiable with respect to optical parameters, reducing memory usage and runtime by up to 6× while enabling the use of more expressive density models that produce better designs.

We demonstrate the effectiveness of IDEAL-IO across diffractive optics, lensless imaging, and snapshot 3D microscopy, showing both computational advantages and design improvements. Our results establish information-theoretic optimization as a practical, scalable strategy for next-generation imaging system design, offering a powerful alternative to traditional approaches without their associated computational burdens.

\section{Background}
Our goal is to optimize a learnable element $\theta$ of an optical system $f(\cdot;\theta)$. For a scene or object $\mathbf{O}$, the system deterministically gives noise-free data $\mathbf{X}=f(\mathbf{O};\theta)$ corrupted by a stochastic noise model $\epsilon$ to yield noisy measurements $\mathbf{Y(\mathbf{{O};\theta})}=\epsilon\left (f(\mathbf{O; \theta})\right )$. We assume access to a dataset of objects $\mathcal{D}_O=\{o_i\}_{i=1,..,N}$ drawn from a distribution $p(\mathbf{O})$ for which we would like to optimize the system. In this section, we describe prior work on how to formulate this optimization problem.

\paragraph{End-to-End Learning.} Previously developed end-to-end learning approaches~\cite{lin_end--end_2020,sitzmann_end--end_2018,sun_end--end_2021} train a neural network $g(\cdot;\phi)$ to perform reconstruction on a dataset of measurements $\mathcal{D}_Y=\left \{y_i\right\}_{i=1,...,N}$ simulated from $\mathcal{D}_O$. These methods use gradient descent-based strategies to jointly optimize 
\begin{equation}
\label{eq:e2e}
\theta^*,\phi^* = \arg \min_{\theta,\phi} \sum_i \mathcal{L}\left[g\left (\epsilon\left (f\left (o_i;\theta \right )\right );\phi\right ),o_i\right ],
\end{equation}
where $\mathcal{L}$ is a loss describing the fidelity of the reconstruction to the original object. While effective, these methods require substantial computational overhead to differentiate Eq.~\ref{eq:e2e} through both reconstruction network $g$ and forward model $f$~\cite{kellman_memory-efficient_2020}.  Moreover, end-to-end optimization can get stuck in local minima and struggle to obtain accurate gradients when reconstructions have not yet converged~\cite{yang_image_2023}. They also yield a design specific to a particular reconstruction algorithm, and this design may become non-ideal if the algorithm is updated or improved.

\paragraph{IDEAL.}
An alternative strategy, Information-Driven Encoder Analysis Learning (IDEAL) was proposed~\cite{anonymous2025information} to avoid differentiation through and sensitivity to the reconstruction algorithm. IDEAL instead maximizes the mutual information $I(\mathbf{O}; \mathbf{Y})$ between the object $\mathbf{O}$ and noisy measurement $\mathbf{Y}$. Because our optical system is deterministic, i.e. $\mathbf{X} = f(\mathbf{O}; \theta)$, this is equivalent to maximizing $I(\mathbf{X}; \mathbf{Y})$, which can be decomposed as
\begin{equation}
\label{eq:mi}
I(\mathbf{X}; \mathbf{Y}) = H(\mathbf{Y}) - H(\mathbf{Y} | \mathbf{X}).
\end{equation}

$H(\mathbf{Y})$ is the entropy of the noisy sensor measurements and captures variability due to both the scene and sensor noise. $H(\mathbf{Y} | \mathbf{X})$ quantifies the uncertainty in $\mathbf{Y}$ due solely to noise, conditioned on the underlying noiseless measurement. The difference of these terms isolates the portion of measurement variability due to the scene, and forms the core objective for information-driven optical design.

The conditional entropy, $H(\bf{Y}|\bf{X})$ can be computed analytically for Poisson and Gaussian noise; these analytical expressions are derived in Supplement Sec. 1. 
Accurately computing $H(\bf{Y})$, on the other hand,
requires access to the true distribution $p(\mathbf{Y})$. In practice, this distribution is approximated by a parameterized model $p_\psi(\mathbf{Y})$, trained on $\mathcal{D}_Y$. This allows us to estimate an upper bound on the true entropy using the cross-entropy over a held-out test set of $M$ noisy measurements:
\begin{align}
\label{eq:h_y}
H(\mathbf{Y}) &\leq \hat{H}(\mathbf{Y}) =\mathbb{E}_{\mathbf{Y}}[-\log p_\psi(\mathbf{Y})]
 \\
&\approx -\frac{1}{M} \sum_{i=1}^M \log p_\psi(\mathbf{y}^{(i)}).
\end{align}

The tightness of the bound in Eq.~\ref{eq:h_y} depends on the expressiveness of the model class $p_\psi$. 
More expressive models—such as PixelCNN~\cite{pixelcnn} or autoregressive transformers~\cite{vaswani_attention_2017}—enable more accurate entropy estimates and, in turn, more informative optimization signals. This approach also allows systematic comparison across density models: lower cross-entropy values indicate better fit to the measurement distribution.
Since this cross-entropy upper bounds the true entropy, it offers a systematic approach for model comparison: models yielding lower cross-entropy values provide more precise distribution estimates. This is used to select the best model for the given scene dataset and optical forward model.

The dependence of Eq.~\ref{eq:h_y} and thus Eq.~\ref{eq:mi} on the \textit{fitted} probability distribution $p_\psi$ introduces significant complexity in the optimization. In the IDEAL setting, we aim to solve $\theta^* = \arg \min_{\theta} -I(\mathbf{X};\mathbf{Y})$. The appropriate gradient descent update at step $t$ is given by:
\begin{align}
\label{eq:ideal_sgd}
\theta_{t+1} &= \theta_{t}-\alpha\frac{\partial I(\mathbf{X};\mathbf{Y})}{\partial \theta}\Bigr\rvert_{\theta = \theta_t} \\
&\approx \theta_t - \alpha\left \{ -\frac{1}{M}\sum_{i=1}^{M}\frac{1}{p_{\psi(\theta_t)}(y^i)}\left [\frac{\partial p}{\partial \psi}\frac{\partial \psi}{\partial \theta}+\frac{\partial p}{\partial y}\frac{\partial y}{\partial \theta}\right ]-\frac{\partial H(\mathbf{Y}|\mathbf{X})}{\partial \theta}\right\}\Biggr\rvert_{\theta = \theta_t} 
\end{align}
where $\alpha$ is the step size and a detailed derivation is given in Supplement Sec. 2. 
Of particular note, this update includes the quantity $\frac{\partial p}{\partial \psi}\frac{\partial \psi}{\partial \theta}$; the mutual information estimate depends on $\theta$ not only through the simulated measurements $y$ but also through the fitted probability distribution parameters $\psi$. In previous work, IDEAL proposed computing gradients in a single step, using a multivariate Gaussian as $p_{\psi}$, estimating its covariance matrix from a batch of measurements, and using the analytic expression for its entropy. The downside of this approach is that it requires differentiating \textit{through} this fitting procedure, which is computationally costly for large covariance matrices on high-dimensional images, and is practically impossible for more expressive parameterizations of $p$, such as neural networks, where the fitting process is typically an iterative procedure.

\section{Methods}
Our contribution is built on an empirical observation that the quantity $\frac{\partial \psi}{\partial\theta}$ is small; i.e., the rate of change of the fitted probability distribution parameters with respect to the learned imaging system parameters is small. We can then ignore the dependence of the mutual information on $\theta$ through the fitted parameters $\psi$. This allows us to dramatically simplify the optimization procedure by ignoring the $\frac{\partial p}{\partial \psi}\frac{\partial \psi}{\partial\theta}$ term and simply refitting $\psi$ outside of the differentiation path to maintain an accurate estimate of $p_\psi$. Our proposed optimization scheme is thus:

\begin{align}
\label{eq:idealio_sgd}
\psi_{t+1} &= \arg \min_{\psi} \mathbb{E}_{\mathbf{Y}}[-\log p_\psi(\mathbf{Y}(\theta_t))]\\
\theta_{t+1} &= \theta_t - \alpha\left \{ -\frac{1}{M}\sum_{i=1}^{M}\frac{1}{p_{\psi_{t+1}}(y^i)}\left [\frac{\partial p}{\partial y}\frac{\partial y}{\partial \theta}\right ]-\frac{\partial H(\mathbf{Y}|\mathbf{X})}{\partial \theta}\right\}\Biggr\rvert_{\theta = \theta_t}. 
\end{align}

\begin{figure}[htbp]
\centering
\includegraphics[width=1.0\linewidth]{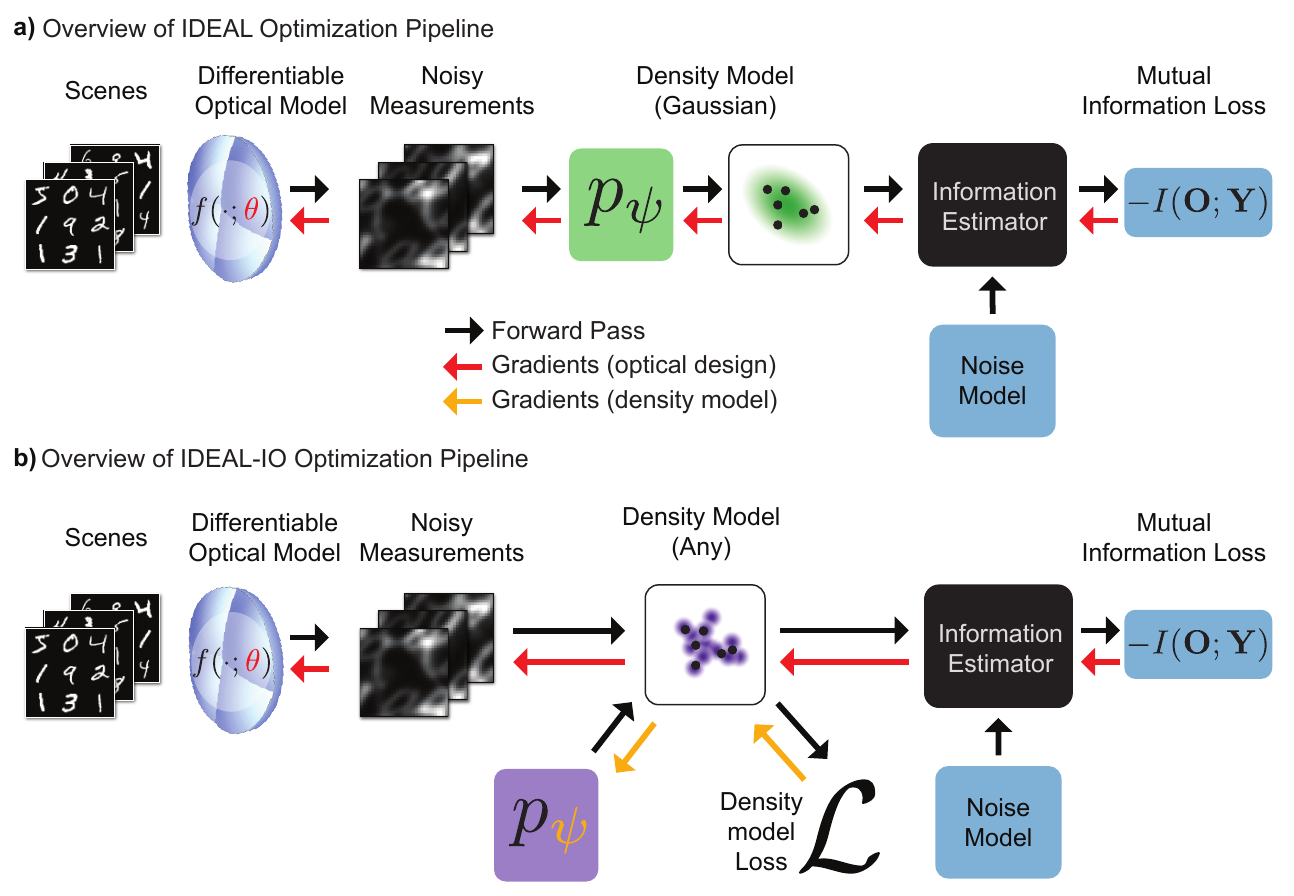}
\caption{\textbf{Comparison of Information-Driven Encoder Analysis Learning (IDEAL) and  IDEAL with Interchanging Optimization (IDEAL-IO) optimization frameworks.}
\textbf{(a)} IDEAL jointly optimizes a differentiable density model and imaging system to maximize mutual information (MI) between noiseless and noisy measurements, $I(\mathbf{X};\mathbf{Y})$. This coupling requires the entire pipeline to be differentiable with respect to the optical parameters, resulting in high memory and compute overhead and limited density model choices.
\textbf{(b)} IDEAL-IO decouples model fitting and optical updates. A density model (e.g., PixelCNN) is fit to noisy measurements with gradients frozen. The fit density model and known noise distribution are then used to estimate MI, which is backpropagated through the imaging system. This alternation substantially lowers computational costs, enabling the use of more expressive models.}
\label{fig:bayer_ideal}
\end{figure}

Intuitively, this scheme alternates between (1) fitting the probabilistic model to measurements generated with the current learned optical element and (2) updating the learned optical element using the most recent probabilistic model. In practice, we implement our method by freezing gradient tracking during the fitting of the probability distribution, and then relying on auto-differentiation to compute the gradient in Eq.~\ref{eq:idealio_sgd}.  To reduce memory usage, we operate on measurement patches rather than full-frame measurements during both model fitting and mutual information estimation. The full optimization pipeline is illustrated in Fig.~\ref{fig:bayer_ideal}. 

This modification dramatically improves the  usability of IDEAL. Because we no longer need to differentiate through the model fitting procedure, we are able to greatly reduce the memory and runtime cost of our algorithm. Further, we no longer have to restrict to parameterizations of $p$ whose fitting process is differentiable, enabling us to use complex, expressive models previously intractable under IDEAL. In particular, we can use neural networks like PixelCNN~\cite{pixelcnn} to model complex, high-dimensional probability distributions that are not Gaussian in nature.

Of note, while the Gaussian model can be fit almost instantaneously, the PixelCNN model requires a significantly longer time to fit (order of minutes). Consequently, when using the PixelCNN model, we opt not to re-fit after each imaging system update but rather every $K$ optical parameter updates, as the loss only needs to provide useful gradient directionality—not an exact estimate of mutual information. For further discussion on PixelCNN re-fit frequency see Supplement Sec. 3.

\section{Results}
We next describe a series of applications showing the advantages of the IDEAL-IO method. First, we show that IDEAL-IO performs comparably to end-to-end design for designing a learned lens array for a snapshot 3D microscopy system while requiring substantially less runtime. Next, we show that the IDEAL-IO alternating update scheme enables faster, more memory-efficient optimization of a large diffractive optical element than standard IDEAL. Finally, we show that IDEAL-IO's compatibility with more expressive probability density parameterizations enables design of a superior point spread function for lensless imaging. We provide an overview of each imaging system within the appropriate subsection, and formal analytical expressions for each system's forward model are available in Supplement Section 9. 

\subsection{Snapshot 3D Microscopy: IDEAL-IO is Faster Than End-to-End Learning}

\begin{figure}[htbp]
  \centering
  \includegraphics[width=0.9\linewidth]{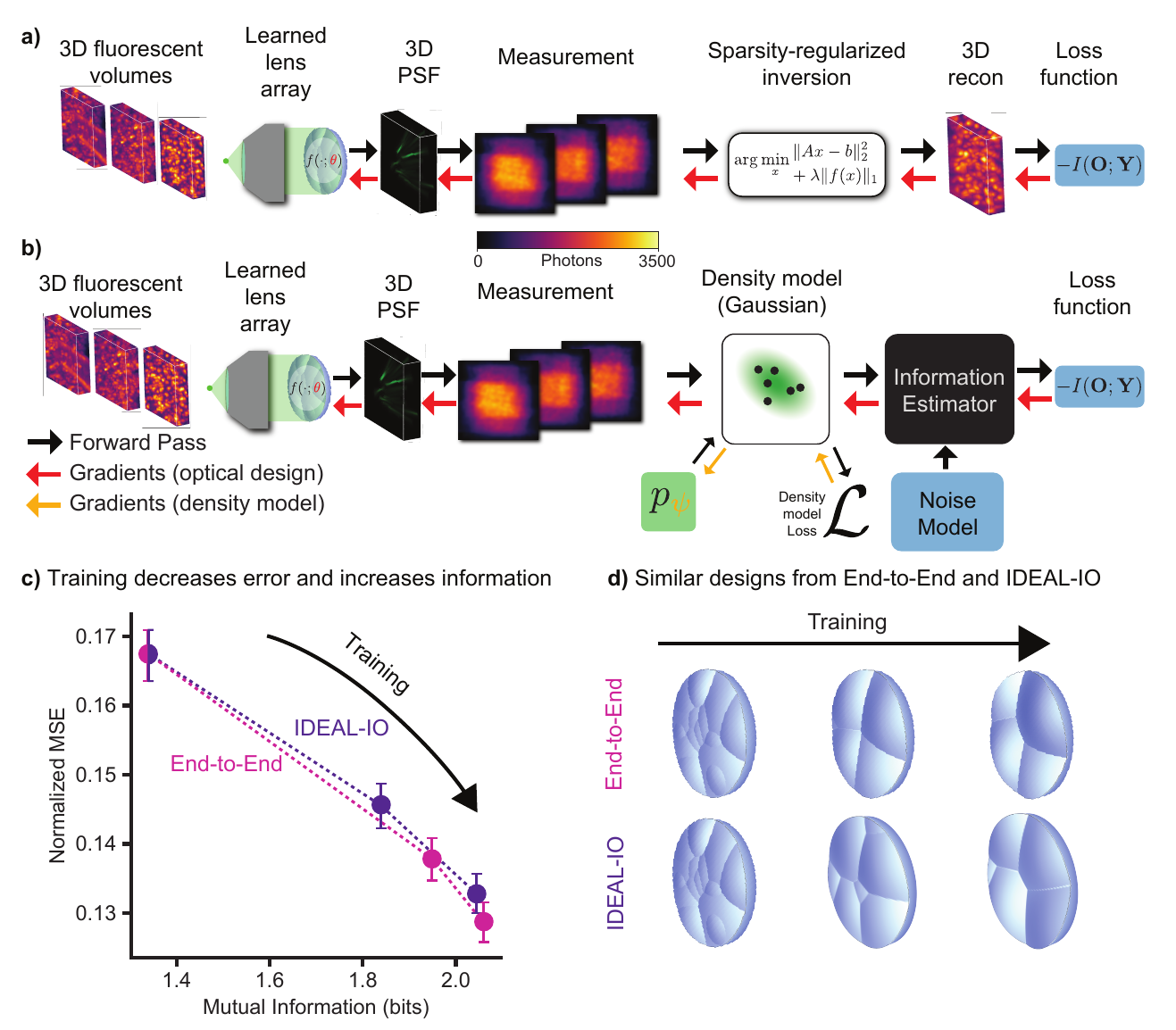}
    \caption{\textbf{Comparison of IDEAL-IO and end-to-end (E2E) design for snapshot 3D microscopy.}
    \textbf{(a)} E2E pipeline. A learned lenslet array encodes a 3D fluorescent volume into a 2D measurement. A neural decoder (FISTA-Net) reconstructs the volume which is fed into a loss function along with the ground truth volume. Gradients are backpropagated through the full pipeline to optimize the optical encoder. 
    \textbf{(b)} IDEAL-IO pipeline. A density model is fit to simulated measurements (with gradient tracking off), then used to estimate mutual information (MI) (with gradient tracking turned on).  The MI estimate is negated and serves as the loss function to drive encoder optimization.
    \textbf{(c)} MI vs. normalized mean squared error (NMSE) at three checkpoints: initialization, mid-training, and convergence for both IDEAL-IO and E2E. MI is estimated using a Gaussian model; NMSE is computed using a separately trained decoder. 
    \textbf{(d)} Lenslet arrays learned by both methods at the checkpoints in \textbf{(c)}. Both methods converge to similar four-lenslet designs that distribute focal points across the volume depth of interest.}
    \label{fig:fourier_diffuserscope}
    \end{figure}

First, we evaluate IDEAL-IO on a simulated snapshot 3D fluorescence microscopy system modeled after the Fourier DiffuserScope~\cite{liu_fourier_2020}. This architecture places a lenslet array in the Fourier plane of a conventional fluorescence microscopy setup to encode a 3D fluorescent volume into a 2D sensor measurement. Then, a neural decoder (FISTA-Net~\cite{xiang_fista-net_2021}) reconstructs the volume.  In prior work, the entire system has been optimized end-to-end (E2E) by minimizing reconstruction error~\cite{Markley2021} through joint optimization of the lenslet array encoder and neural decoder. 
While effective, E2E optimization is computationally intensive. We hypothesize that its success stems in part from implicitly increasing the MI between the scene and measurement. If true, directly maximizing MI should yield comparable encoder designs without the overhead of decoder training during optimization.

To test this, we train the E2E system  on fluorescense microscopy images of mouse lungs and evaluate encoder quality at three checkpoints: initialization, midway through training, and convergence. At each point, we compute MI using a Gaussian density estimator and assess reconstruction fidelity via normalized mean squared error (NMSE) by training a reconstruction network with the optic fixed. As shown in Fig.~\ref{fig:fourier_diffuserscope}c, MI increases monotonically as NMSE decreases, supporting the hypothesis that E2E optimization indirectly maximizes MI.

We then apply IDEAL-IO to the same design task. Despite never training a decoder during optimization, IDEAL-IO produces an encoder with comparable MI and NMSE to the E2E-trained system. The learned encoder exhibits a similar lenslet configuration (Fig.~\ref{fig:fourier_diffuserscope}d), with focal points distributed across depth.

IDEAL-IO converges in $\sim$25 minutes on a single RTX A6000 GPU—roughly 4$\times$ faster than the $\sim$2 hour E2E baseline. Unlike E2E, which required memory-saving techniques such as gradient checkpointing~\cite{kellman_memory-efficient_2020}, IDEAL-IO ran without special accommodations. These results demonstrate that MI is not only a valid proxy for reconstruction fidelity, but also a practical and scalable objective for optical encoder design.

\subsection{Diffractive Optical Elements: IDEAL-IO is Faster, More Memory-Efficient than IDEAL}
\label{sec:doe}
\begin{figure}[!hbp]
  \centering
  \includegraphics[width=0.85\linewidth]{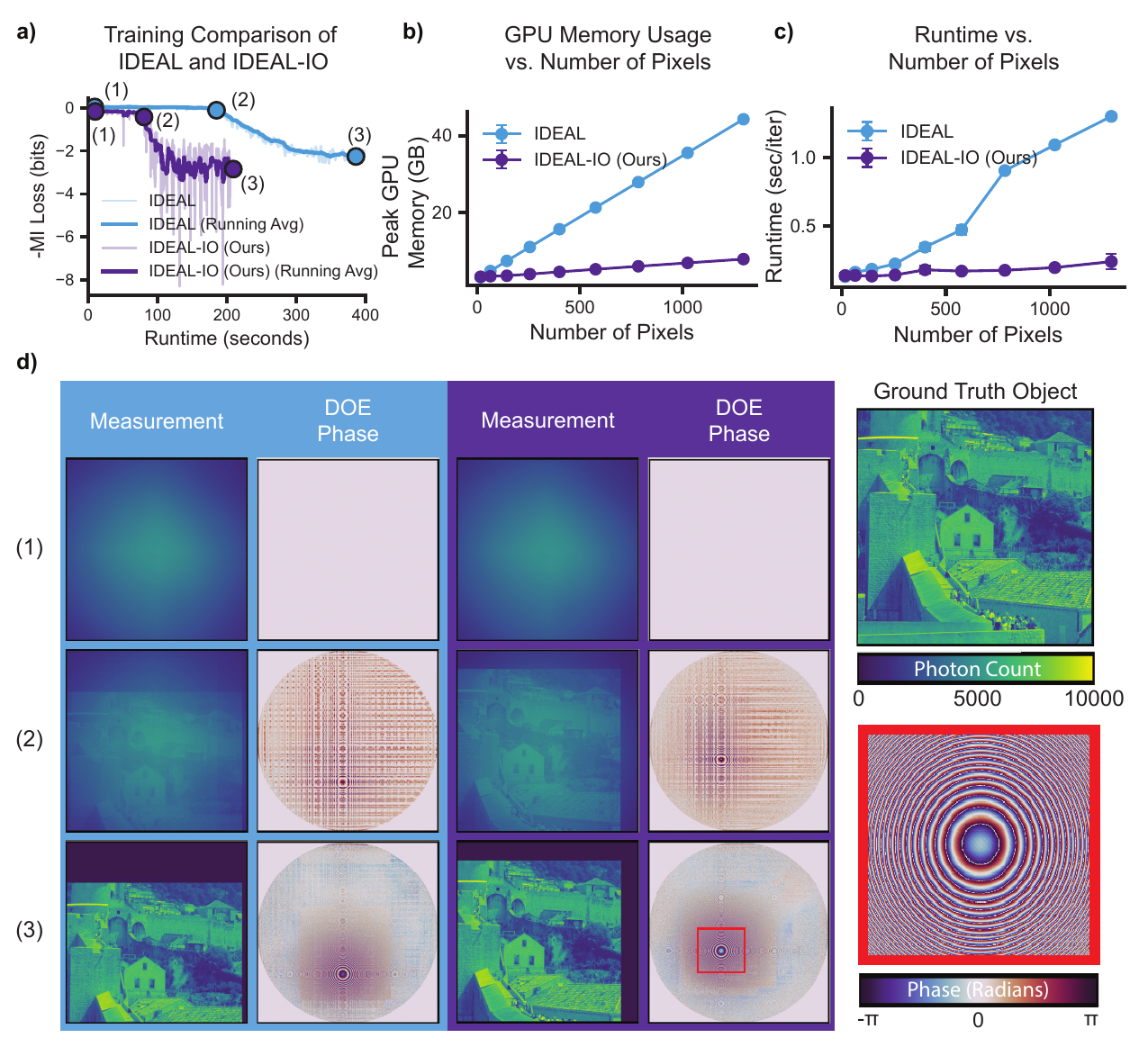}
    \caption{\textbf{Scaling behavior and learned designs for diffractive optical element (DOE) optimization with IDEAL and IDEAL-IO.}
    \textbf{(a)} Optimization of a DOE for single-plane imaging using IDEAL (blue) and IDEAL-IO (ours, purple).  Training loss curves are shown for both methods.  
    \textbf{(b)} Runtime per optimization step vs. number of pixels in a patch used for mutual information estimation. IDEAL runtime increases rapidly with patch size; IDEAL-IO remains relatively constant. Error bars denote standard deviation across the 10 repeated trials; when they are not visible, the error is smaller than the data marker.
    \textbf{(c)} Peak GPU memory usage vs. patch size. IDEAL memory usage grows steeply; IDEAL-IO scales more favorably.
    \textbf{(d)} Visualizations of the DOE phase profile and resulting measurement from (1) initialization, (2) an intermediate training step, and (3) convergence, for both methods.  Despite no structural priors, both IDEAL and IDEAL-IO converge to Fresnel-like designs.
  }
  \label{fig:doe_ideal}
\end{figure}

To compare GPU memory requirements and runtime of our method, IDEAL-IO, to the original implementation of IDEAL, we apply it to a optical design task with a large number of learnable parameters: optimizing a pixelwise height map for a diffractive optical element (DOE) used in single-plane imaging \cite{sitzmann_end--end_2018}.  This problem involves optimizing a 2496 $\times$ 2496 pixel height map resulting in $\sim6.2$ million learnable optical parameters.  We modify the code from~\cite{sitzmann_end--end_2018} and use the dataset from~\cite{jegou_hamming_2008} to perform this optimization.

This DOE is modeled as a flat surface with a spatially-varying phase profile, where each pixel in the height map imparts a phase delay to the incoming wavefront according to
\[
\phi(x, y) = \frac{2\pi}{\lambda} \, \Delta n \, h(x, y),
\]
where \(h(x, y)\) is the pixel height, \(\lambda = 650\,\text{nm}\) is the design wavelength, and \(\Delta n = 1.4599 - 1.0\) is the refractive index contrast between the DOE material and air. The refractive index is assumed to be uniform across the surface.
To simulate image formation, we use angular spectrum propagation, a standard wave optics technique that models wave propagation~\cite{goodman2005fourieroptics}. To generate sensor measurements, we convolve this point spread function with the input object and add noise based on the Gaussian approximation of Poisson noise.

Despite the large number of learnable parameters, IDEAL-IO converges in $\sim$120 seconds on a single Nvidia RTX A6000 GPU using only 7.89 GB; in contrast, IDEAL converges in $\sim$360 seconds using 44.23 GB.  Both optimizations were carried out using patches of size 36 $\times$ 36 pixels and used 6480 patches to fit the Gaussian model.

For this test case, both IDEAL and IDEAL-IO converge to  something close to the expected design -- a Fresnel zone plate -- despite no explicit structural priors being imposed.  A Fresnel zone plate is the binary diffractive optical equivalent of a traditional lens, which should be optimal for 2D imaging of dense, natural scenes~\cite{leylalensless}. Figure~\ref{fig:doe_ideal}a shows the training loss curves along with visualizations of the learned DOEs and corresponding measurements at selected iterations. Both methods show similar convergence, though IDEAL-IO's curve appears noisier due to its smaller test set and mutual information's sensitivity to outlier patches. 

To quantify memory and runtime differences between methods, we ran optimizations using patch sizes ranging from $4 \times 4$ to $36 \times 36$ pixels. For each patch size, we performed 50 optimization steps using a training set of measurements equal to 5$\times$ the number of pixels in the patch (e.g., 80 measurements for a $4 \times 4$ patch). This was repeated 10 times per patch size for each method. In Fig.~\ref{fig:doe_ideal}b--c, we plot the mean peak GPU memory usage and the mean runtime per optimization step.  IDEAL-IO provides a substantial decrease in both GPU memory usage and runtime compared to IDEAL. 

To better quantify memory scaling, we fit a linear regression model between the number of pixels in the patch and peak GPU memory usage. We report the number of additional patch pixels required to increase GPU usage by 1 GB, computed as the inverse slope of the fit. Similarly, we fit a linear model to runtime per step as a function of pixel count, and report the runtime increase per 100 additional pixels. These results are presented in Table~\ref{tab:scaling} and demonstrate that our method preserves the performance benefits of mutual information–based design while dramatically improving scalability and runtime.

\begin{table}[tbp]
  \centering
  \caption{\textbf{Scaling of runtime and GPU memory with patch size in mutual information estimation.} We report the (1) number of additional patch pixels required to increase GPU memory usage by 1 GB, and (2) increase in runtime per optimization step for every additional 100 patch pixels.}
  \label{tab:scaling}
  \begin{tabular}{lcc}
    \toprule
    \textbf{Method} & \textbf{Patch pixels per +1 GB memory} & \textbf{Runtime increase per +100 pixels (s)} \\
    \midrule
    IDEAL & \(31 \pm 0\) & \(0.098 \pm 0.007\) \\
    IDEAL-IO (Ours) & \(274 \pm 4\) & \(0.007 \pm 0.001\) \\
    \bottomrule
  \end{tabular}
\end{table}

\subsection{Lensless Imaging: IDEAL-IO Enables Use of Expressive Probability Density Models}

\begin{figure}[htbp]
  \centering
  \includegraphics[width=0.9\linewidth]{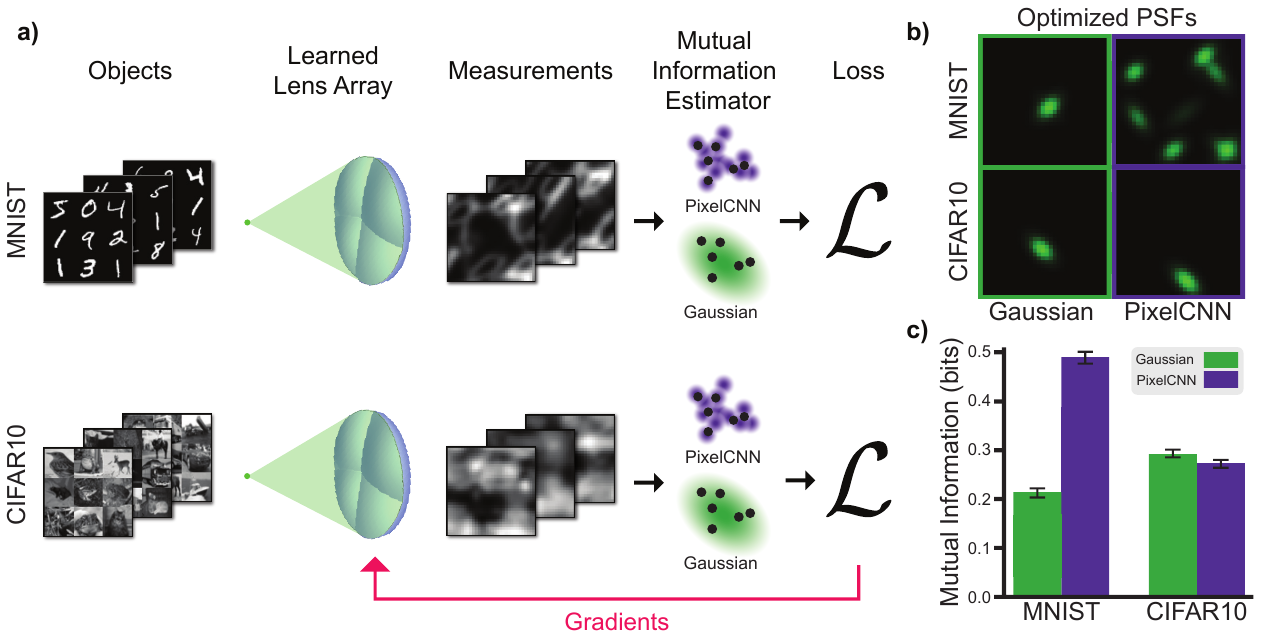}
  \caption{\textbf{IDEAL-IO improves lensless imaging design by using expressive mutual information estimators.}
    (a) IDEAL-IO is used to optimize the PSF of a lenslet array by maximizing mutual information (MI) between scenes and noisy measurements. We compare two density models used in the MI estimator: a Gaussian model and a more expressive PixelCNN. CIFAR10 measurements are approximately Gaussian, while MNIST measurements are sparse and non-Gaussian. 
    (b) Optimized PSFs differ significantly for MNIST depending on the density model chosen for the MI estimator, reflecting the impact of model expressivity. For CIFAR10, both models yield similar PSFs.
    (c) PixelCNN-based designs achieve higher MI on MNIST test data, indicating better alignment with the true measurement distribution. For CIFAR10, both estimators perform similarly due to the near-Gaussian nature of the measurements.
  }
  \label{fig:gaussian_vs_pixelcnn}
\end{figure}

To assess how density model expressivity affects optical design, we revisit a key trade-off identified in prior work~\cite{anonymous2025information}: simple models like Gaussians enable fast optimization but may fail when measurements exhibit non-Gaussian structure. We use IDEAL-IO's compatibility with highly-expressive probabilistic models to investigate whether using a more expressive model during optimization leads to better designs, a comparison not possible with prior work.

We test this question with a simulated lensless imaging system in which a phase mask is placed a short distance in front of the sensor. With this configuration, lensless imagers achieve compact form factors and as a consequence have large distributed PSFs, in contrast to the single-spot PSF that is optimal for the Fresnel-lens in Sec.~\ref{sec:doe}, where the phase mask to sensor distance was larger. Based on prior work in lensless imaging design, a phase mask comprised of a random array of lenslets trades off multiplexing and signal-to-noise ratio well~\cite{kabuli2023high}. Hence, we choose to model the phase mask as an array of lenslets, with its PSF modeled as a sum of 2D Gaussians, each representing the focal spot of an individual lenslet. The mean of each Gaussian determines the focal point location on the sensor, while the covariance controls its shape and orientation. Both the means and covariances are learned during optimization.  See Supplement Sec. 7 for the initialized PSF.

To evaluate the role of density model expressivity, we optimize this system using mutual information estimated with a density model of a simple multivariate Gaussian or a more expressive autoregressive PixelCNN and a Poisson noise model. We test both approaches on CIFAR10~\cite{CIFAR10}, where natural image statistics yield approximately Gaussian measurements, and MNIST~\cite{MNIST}, where sparse digit structures produce highly non-Gaussian, often bimodal, measurement distributions.
We fit both density models using 16,024 patches of size of 16 $\times$ 16 pixels.  The PixelCNN density model was refit after every 50 optical optimization steps in order to reduce runtime, while the Gaussian model was fit after every optical update due to the minimal computational cost of refitting.  The learned estimators were evaluated on 1602 patches to calculate the mutual information loss.

For the CIFAR10 dataset, both estimators converge to similar optical designs with nearly identical loss curves. However, for MNIST, the two estimators yield distinct designs, revealing that model expressivity significantly affects optimization when the true measurement distribution deviates from Gaussian assumptions (Fig.~\ref{fig:gaussian_vs_pixelcnn}).

To evaluate design quality, we compute MI on held-out test data using a PixelCNN estimator. For CIFAR10, both designs perform similarly (0.283 vs. 0.264 bits/pixel), but for MNIST, the PixelCNN-optimized design significantly outperforms the Gaussian one (0.473 vs. 0.208 bits/pixel), confirming that expressive models yield better designs for non-Gaussian data.  These results align with ~\cite{leylalensless}.

\section{Discussion}
In this work, we introduced and rigorously-tested a low computational cost method for imaging system design that builds on the IDEAL framework~\cite{anonymous2025information} . Like IDEAL, our approach optimizes optical parameters by maximizing the mutual information between noiseless and noisy sensor measurements. Our key contribution is a decoupled, two-stage optimization procedure that eliminates the need for the density model  fitting to be differentiable with respect to the optical parameters. By separating density model fitting from differentiable mutual information estimation, our method supports more expressive models, reduces memory usage, and improves runtime, with maintained or improved performance.  We demonstrated the effectiveness of the proposed method across three distinct computational imaging problems: diffractive optical imaging, lensless imaging, and snaphot 3D microscopy.

A key constraint in our framework is the reliance on patch-based entropy estimation, which creates a trade-off between model fidelity and memory efficiency. When using density models like multivariate Gaussians, the number of parameters increases quadratically with patch size. As the dimensionality grows, more samples are needed to fit the model reliably, which increases the computational burden.
At the same time, larger patches reduce the number of unique samples that can be extracted from each measurement. This lowers the statistical diversity of the training set and forces the simulation of more measurements to maintain performance. The result is a significant increase in both memory use and runtime, and may limit modeling capacity in some cases. Future work could explore multiscale or hierarchical approaches to relax this limitation without adding excessive computational overhead.

Mutual information offers a strong foundation for reconstruction-based imaging, as preserving scene information in the system measurement theoretically enables perfect reconstruction.  However, mutual information may not always align with specialized downstream tasks. In tasks where most information may be concentrated in a small subset of the scene, such as classification or detection, maximizing total information content might lead to designs that prioritize high-entropy regions that are irrelevant to the task. Extending our framework to support task-specific variants of mutual information, such as conditional or class-aware formulations, could improve performance in these settings.

Finally, the computational efficiency of our method enables large-scale exploration of design spaces. Multiple initializations can be evaluated quickly, revealing whether optimization consistently converges to similar designs or is sensitive to initialization. This flexibility also facilitates perturbation analyses, which can quantify sensitivity to noise, model error, or manufacturing tolerances—critical factors in real-world deployment. 

By directly maximizing the information content of measurements, IDEAL-IO produces high-performance optical systems without the computational burden of jointly training reconstruction networks while designing the optics. This approach bridges the gap between theoretical guarantees and practical system design, providing both performance benefits and computational efficiency. Our framework serves as a scalable foundation for next-generation imaging system design. Future work could further broaden its applicability by incorporating task-specific constraints, multichannel mutual information estimators, and theoretical analyses of the connections between mutual information and downstream task performance.

\begingroup
\small
\bibliographystyle{unsrt}
\bibliography{references,references-2}
\endgroup

\end{document}